\documentclass[pra,twocolumn,showpacs,,superscriptaddress]{revtex4}

\usepackage{graphicx,epsfig}
\usepackage{times,amsmath,amssymb,latexsym}
\usepackage{rotating}
\usepackage{stmaryrd}

\newcommand{\bs}{\boldsymbol}
\begin{document}

\title[Few-particle systems: An analysis of some strongly correlated states]{Few-particle systems: An analysis of some strongly correlated states}

\author{N. Barber\'an}

\address{Departament de F\'isica Qu\`antica i Astrof\'isica, Facultat de F\'isica, Universitat de Barcelona,
E-08028 Barcelona, Spain}

\author{J. Taron}

\address{Departament de F\'isica Qu\`antica i Astrof\'isica, Facultat de F\'isica, Universitat de Barcelona,
E-08028 Barcelona, Spain}
\address{Institut de Ci\`encies del Cosmos, E-08028 Barcelona, Spain}

%\title{Few particle systems. An analysis of some strongly correlated states.}
%\author{N. Barber\'an$^{1}$, and J. Taron$^{1,2}$}
%\affiliation{(1) Departament de F\'isica Qu\`antica i Astrof\'isica, Facultat de F\'isica, Universitat de Barcelona,
%E-08028 Barcelona, Spain\\
%(2) Institut de Ci\`encies del Cosmos, E-08028 Barcelona, Spain\\ 

%\address{$^1$ Departament de F\'isica Qu\`antica i Astrof\'isica, Facultat de F\'isica, Universitat de Barcelona,
%E-08028 Barcelona, Spain}
%\address{$^2$ Institut de Ci\`encies del Cosmos, E-08028 Barcelona, Spain} 

\begin{abstract}
The analysis of the quantum Hall response of a small system of interacting  ultracold bosonic atoms through the variation of its Hall resistivity against the applied gauge magnetic field, provides a powerful method to unmask its strongly correlated states in a quite exhaustive way. Within a fixed range of values of the magnetic field in the lowest Landau level regime, where the resistivity displays two successive plateaux, we identify the implied states as the Pfaffian and the state with filling factor $\nu=2/3$ in the thermodynamic limit. We fix the conditions to have good observability.

\end{abstract}

\date{\today }
\maketitle

\section{ Introduction}

For a long time, one of the main goals in the study of the many-body interacting systems has been the localization and description of their stable states. In this research area, a special role is played by the strongly correlated states of charges under large magnetic fields \cite{dem,lew}. For some of these states, the analytical expression of their wave functions is available, as in the case of the Laughlin \cite{lau} and the Pfaffian \cite{moo,rea}. The Laughlin is the exact solution of a system under $2$-body contact interaction and the Pfaffian is the exact solution for charges under $3$-body interaction \cite{gre}. However, in general, the difficulty of the analysis has led to the development of quantum simulation as one of the most fruitful ways to deal with the problem \cite{boa,cir,blo}. In the quantum simulation, charges are replaced by neutral cold atoms and real fields by artificial gauge fields. 

\medskip

The simulation of the transport equation given by    
\begin{equation}
j_y\,\,=\,\,\sigma_{yx}\,\,E_x\,\,,
\end{equation}
provide the possibility to analyze the correlated states associated with the resistivity plateaux in a typical experimental outcome showing $\rho_{yx}$ ($\rho_{yx}=\sigma_{yx}/(|\sigma_{xx}|^2+|\sigma_{yx}|^2$)) as a function of the magnetic field \cite{yos}. For a two dimensional system, in Eq.(1), $j_y$ is the mean current density along the $Y$ direction, $\sigma_{yx}$ is the Hall conductivity ($\sigma_{xx}$ comes from $j_x\, = \sigma_{xx}\,E_x$),  and $E_x$ is a time dependent periodic perturbation that simulates the electric field applied in the $X$ direction. Le Blanc {\it et al.} \cite{leb} reported for the first time the experimental signature of quantum Hall effect in a large system of bosonic atoms.  

\medskip

Our ansatz assumes a cloud of interacting ultracold bosonic atoms trapped by a parabolic potential in the XY plane, rotating around the Z-axis. The rotation frequency $\Omega$ simulates the magnetic field \cite{fet}. Along the X-axis we consider a fixed impurity which becomes a necessary condition to visualize the plateaux, as will be discussed in Section III. We perform exact diagonalization in the lowest Landau level regime for strong magnetic fields and describe the system from the rotating frame of reference. Within this regime,  no mean field theories can be applied. The system cannot be characterized by a unique function that plays the role of an order parameter.

\medskip

 Our analysis can only be performed for a relatively small number of particles. However, as it has been stressed \cite{coo1}, small systems provide a good option of experimental access to the region where vortex liquid states appear. For these small systems, indeed, cases like the Laughlin or the Pafaffian type states preserve its full meaning.  

\medskip

Closely following our previous work \cite{bar} for $N=4$ particles,  first we focus on the Laughlin state and analyze the origin of the plateau of $\rho_{yx}$ in the region  where the expectation value of the ground state angular momentum ($L_{GS}$) lies around $12$. In the circular symmetric system, $L_{GS}=N(N-1)$. Next we extend our analysis to $N=5$. In this case, a huge time consuming effort would be required to obtain a Laughlin type state (at around $L_{GS}=20$). However, despite the effort, the expectations of finding new physics beyond what was obtained already for $N=4$ are very low. Aside from that, the presence of two pseudo-plateaux at lower values of $\Omega$ brings the opportunity to identify the associated states using different tools. E.g., the overlap between the exact solution and different analytical expressions,  or the analysis of the edge excitations. We found that the first plateau is related to the  Pfaffian \cite{moo} state and the second one is well identified with the state of filling factor $\nu=2/3$. The great interest on the Pfaffian is justified by the unique properties of its excitations \cite{ron,jul1} that make it attractive in the context of topological quantum computation. 

\medskip

The identification of liquid vortex states was already considered in previous works \cite{coo1,coo2,wil,reg,caz1,caz2,dag}, some of them  on the torus or spherical geometries.

\medskip

Of  special importance is the analysis of the edge excitations. As stressed by Wen \cite{wen1,wen2,wen3}, the topological order of the fractional quantum Hall (FQH) states is reflected in the properties of their edge excitations. The topological order is a characteristic that classifies the FQH states in a unique way.  

\medskip

We summarize our main results in two points: On the one hand, we analyze the reason why around certain values of $\Omega$ and only at them, say $\Omega_i$, a plateau of the resistivity can be displayed \cite{bar},  localizing all the quantum liquid stable states of definite topology. These states are all the existing ones within the range of $\Omega$-values under study. On the other hand, we identify the implied correlated states in each plateau, using different tools and specially, the analysis of their edge excitations \cite{caz1,caz2}. 

\medskip

In addition, another remarkable result is the direct observation of the ordered pattern in the density of the GS, generated by $3$-body contact interaction. This allows us to give an interpretation of the Pfaffian state. Usually, the spatial correlation of the atoms is hidden in a circular symmetric ground state and the analysis of the two-body pair correlation function ($\,\rho^{(2)}(\vec{r}_0,\vec{r}$)\,)  is necessary to uncover it. One atom is fixed at a given position $\,\vec{r}_0\,$ and the probability of finding the other ones around it is correlated. For example, in the case of the Laughlin state, a triangle of peaks of density is obtained for $N=3$, a square for N=4, etc. However, the existence of an impurity plays a similar role in the density, the position of the impurity breaks the circular symmetry and the ordered pattern is explicit already in the density. But distinctly, the number of peaks directly on the density in the case of 3-body interaction is not equal to the number of particles. Our possible explanation is given in Section V.   

\medskip

Our paper is organized as follows: In Section II we present the model. In Section III  we analyze the origin of the plateaux that emerge in the resistivity around particular values of the magnetic field. In Section IV we analyze the topological order of the implied states at each $\Omega_i$ through their edge excitations. In Section V we show the numerical results and discuss their interpretation. Finally in Section VI we present our conclusions.

\medskip

\medskip

\begin{figure}[tbp]
	\centering
		\includegraphics*[width=6.5cm]{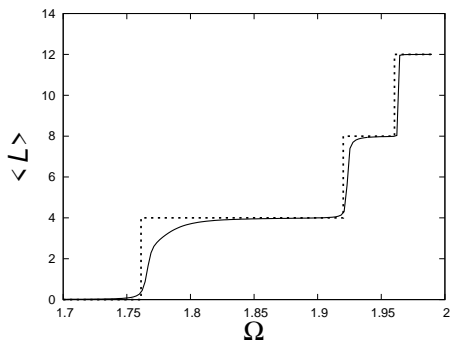}
\caption{Variation of the expectation value of the angular momentum as a function of the rotation frequency $\Omega\,$ in units of $\omega_{\perp}/2$. We consider  $N=4\,$, $\,g_2=1\,$, $\,$ $ \,\gamma=0.1\,$ and $\,\bs a=(1,0)$ in our unit of length (see text). The step-wise dashed line refers to the symmetric case ($\gamma=0$).} 
\end{figure}

\begin{figure}[tbp]
	\centering
		\includegraphics*[width=6.5cm]{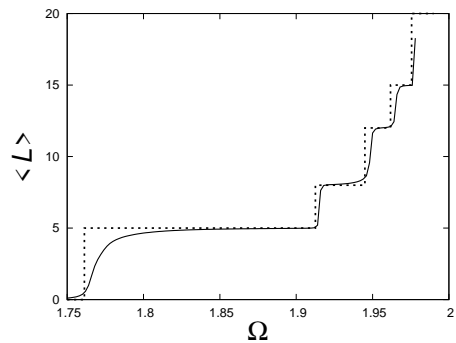}
\caption{ The same as Fig.1. for $N=5$.}  
\end{figure}

\section{ Model}

\begin{figure}[tbp]
	\centering
	\includegraphics*[width=\columnwidth]{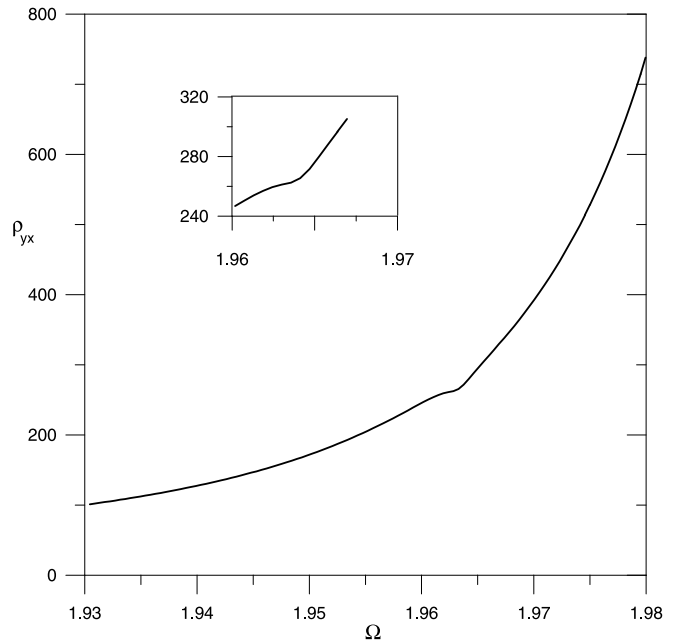}

\caption{ Hall resistivity as a function of $\Omega$. We considered $N=4$, $a=(1.0,0)$, $\gamma=0.1$ and $g_2=1$. The small plateau is approximately at $\Omega=1.963$. We considered our units of length, energy and frequency (see text).} 
\end{figure}

\begin{figure}[tbp]
	\centering
\includegraphics*[width=7.5cm]{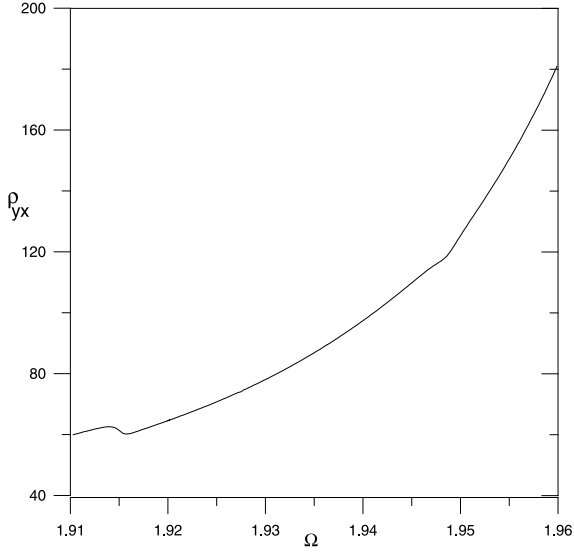}

		\caption{ The same as Fig.3 for $N=5$,  $a=(0.6,0)$, $\gamma=0.1$ and $g_2=1$ . The small plateaux are localized at approximately $\Omega=1.915$ and $\Omega=1.947$, no $3$-body interaction is considered.} 
\end{figure}

\medskip

We closely follow the model presented in Ref \cite{bar} with the addition of a $3$-body contact interaction term in the Hamiltonian.

\medskip

We consider a system of $N$ one-component bosonic atoms of mass $M$ confined on the $XY$-plane. The cloud
is trapped by a rotating parabolic potential of frequency $\omega_{\perp}$ and
rotation $\Omega$ along the $Z$-axis. In the rotating reference frame 
the Hamiltonian reads
\begin{equation}
\hat{H}(t)\,=\, \hat{H}_{sp}\,\,+\,\,\hat{H}_{int}\,\,+\,\,\hat{H}_{pert}(t),
\end{equation}
where the single-particle (sp)  part is given by

\begin{eqnarray}
\hat{H}_{\rm{sp}} & = & \sum_{i=1}^N\bigg[\frac{1}{2M}(\hat{\bs p}+\hat{\bs A} )^2 + \frac{1}{2}M\left( \omega_{\perp}^2-\frac{(B^*)^2}{4M^2}\right)\hat{\bf r}^2 
\nonumber
\\
& - & \gamma \frac{\hbar^2}{M}\delta^{(2)}(\hat{\bs r}- \bs a)\bigg]_i ,
\end{eqnarray}
with
\begin{equation}
\hat{A}_x=\frac{B^*}{2}\hat{y}\,\,\,,\,\,\,\hat{A}_y=-\frac{B^*}{2}\hat{x}\,,
\end{equation}
where $\,\,\bs r=(x,y)\,\,$ and $B^*=2M\Omega$.
The last term is due to the presence of an  impurity modeled by a Dirac delta function. The dimensionless parameter $\gamma$ measures its strength and $\bs a$ localizes it on the $XY$ plane.

\medskip 

The atomic interaction is modeled  by a $2D$ contact potential given by,

\begin{eqnarray}
\hat{H}_{\rm{int}} & = & \frac{\hbar^2 }{M}g_2\sum_{i<j} \delta^{(2)}(\hat{\bs r}_i-\hat{\bs r}_j) 
\nonumber
\\
& + & \frac{\hbar^2}{M}\lambda_{\perp}^2 g_3\sum_{i<j<k}\delta^{(2)}(\hat{\bs r}_i-\hat{\bs r}_j)\delta^{(2)}(\hat{\bs r}_i-\hat{\bs r}_k)\,,
\end{eqnarray}
where $g_2$ and $g_3$ are the dimensionless coupling parameters that give the strength of the $2$ and $3$-body interactions respectively and $\lambda_{\perp}^2\,=\,\frac{\hbar}{M\omega_{\perp}}$. 

\medskip

In the LLL regime without impurities, the kinetic part of the Hamiltonian reads, 
 
\begin{equation}
\hat{H}_{\rm{kin}}=\hbar(\omega_{\perp}-\Omega)\,\hat{L}\,+ \hat{N}\hbar\omega_{\perp}\,.
\end{equation}
The sp solutions with well defined angular momentum $m$ are the Fock-Darwin (FD) functions \cite{jac}, given by $\,\phi_m(\theta,r) \,=\,\frac{e^{im\theta}}{\sqrt{\pi m!}} \,e^{-r^2/2}\,
r^m\,$ where we consider $\lambda _{\perp}$ as the unit of length.

\medskip
Once $\hat{H}_0\,=\,\hat{H}_{sp}\,+\,\hat{H}_{int}\,\,$ is solved, we proceed to diagonalize the one-body density matrix given by,

\begin{equation}
\hat{\rho}^{(1)} (\bs r,\bs r')\,=\,\langle \hat{\Psi}^{\dag}(\bs r) \hat{\Psi}(\bs r')\rangle\,,
\end{equation}
where the expectation value is calculated at the GS and $\hat{\Psi}(\bs r)$ is the field operator. The eigenfunctions are the one-body $\,\it natural\,\, orbitals\,$ $\,{\psi_i}\,$, linear combinations of the FD functions, i.e. $\,\psi_i= \sum_{0}^{m_{max}} p_m^i \phi_m$. The eigenvalues are their occupations $\,{n_i}\,$, $i=1,..,i_{max}$. Notice that $m$ is angular momentum while $i$ is an index that labels the eigenstates, namely $i_{max}= m_{max}+1\,$,  $m_{max}$ is  varied until convergence of the results.

If some of the natural orbitals are localized at the impurity, then we are able to distinguish between two type of states: localized and extended, a crucial condition necessary to understand the mechanism that produces a plateau as will be explained in section III.

\medskip

For the effective periodic electric field we consider

\begin{equation}
\hat{H}_{pert}(t)=-\lambda \frac{\hbar^2}{M\lambda_{\perp}^3}(\sum_{i=1}^N \hat{x}_i) \xi(t) sin(\omega t)\,\,\,\,\equiv \,\,\,\,\sum_{i=1}^N E_x(t) \hat{x}_i
\end{equation}
where $\lambda$ is the dimensionless parameter that gives the intensity of the perturbation which we assume small and 

\begin{equation}
\xi(t)=1-exp[-(t/\sigma)^2]\,\,.
\end{equation}
From now on we consider $M=1/2$
and $\hbar=1$ and choose $\,\lambda_{\perp}=\sqrt{\frac{\hbar}{M\omega_{\perp}}}=\sqrt{2/\omega_{\perp}}\,$, $\,\hbar\omega_{\perp}/2\,$ and $\,\omega_{\perp}/2\,$ as units of length, energy, and frequency, respectively.
With our unit of length, $\,\omega_{\perp}=2$. 

\medskip

Finally, to identify the Hall conductivity $\sigma_{yx}$ from Eq.(1),
we analyze the time evolution of the expectation value of the current operator, $\langle \Psi(t)|\hat{j}_y|\Psi(t)\rangle$ where $\Psi(t)$ is the many-body wave function of the system, until   
the stationary regime is reached.

\begin{figure}[tbp]
	\centering
		\includegraphics*[width=5.5cm,angle=270]{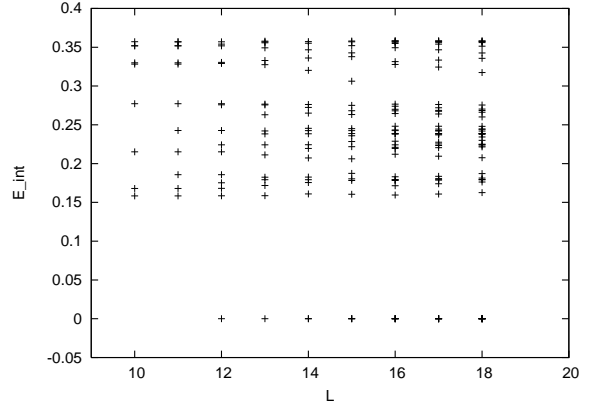}
\caption{ Spectrum around the Laughlin state ($L_{GS}=12$) for $N=4$, The gap at $L=12$ is equal to $0.1585$ in units of $\hbar \omega_{\perp}/2$. The bottom states at $L=13,...,18$ are degenerated. $\Omega\,=\,1.97$ has been considered.} 
\end{figure}

\begin{figure}[tbp]
	\centering
		\includegraphics*[width=5.5cm,angle=270]{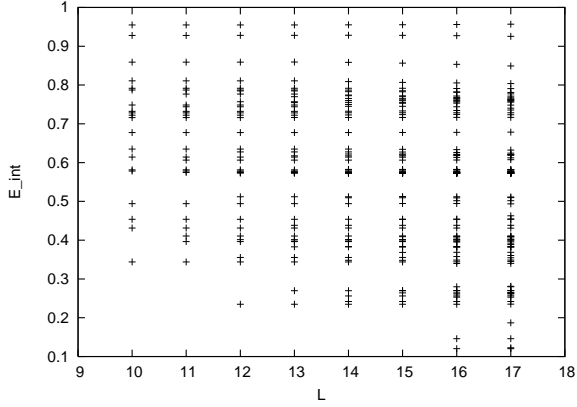}
\caption{ Spectrum around the $\nu=2/3$  state ($L_{GS}=12$) at the second plateau of Fig.4 for $N=5$. The gap at $L=12$ is equal to $0.1091$ in our units. $\Omega\,=\,1.947$ has been considered.} 
\end{figure}

\begin{figure}[tbp]
	\centering
		\includegraphics*[width=6.5cm]{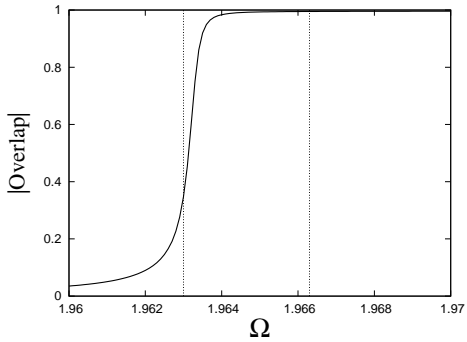}
\caption{ Absolute value of the overlap between the exact solution and the analytical Laughlin expression as a function of $\Omega$. $N=4$, $g_2=1$, $\gamma=0.1$ and $\bs a=(0.6,0)$ is considered. Vertical lines mark the frontiers of the plateau shown in Fig.3, from $\Omega=1.963$ to $1.9663$.} 
\end{figure}

\begin{figure}[tbp]
	\centering
		\includegraphics*[width=6.5cm]{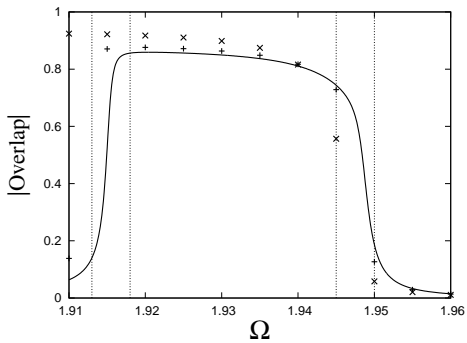}
\caption{ The same as Fig.7 for $N=5$. The analytical function considered on the overlap is the Pfaffian state. The significant values appear over the step $\langle L \rangle = 8$. The full line corresponds to $g_3=0$, the plus symbols to $g_3=1$ and the crosses to $g_3=5$. For all of them, $g_2=1$. We consider $\bs a=(0.6,0)$ and $\gamma=0.1$. The vertical lines mark the frontiers of the first (form $1.913$ to $1.918$) and the second plateau (from $1.945$ to $1.95$) respectively. Notice that in general, as the interaction grows, there is a whole shift to the left of the critical values of $\Omega$ where $L$ jumps.    
} 
\end{figure}

\begin{figure}[tbp]
	\centering
		\includegraphics*[width=7.5cm]{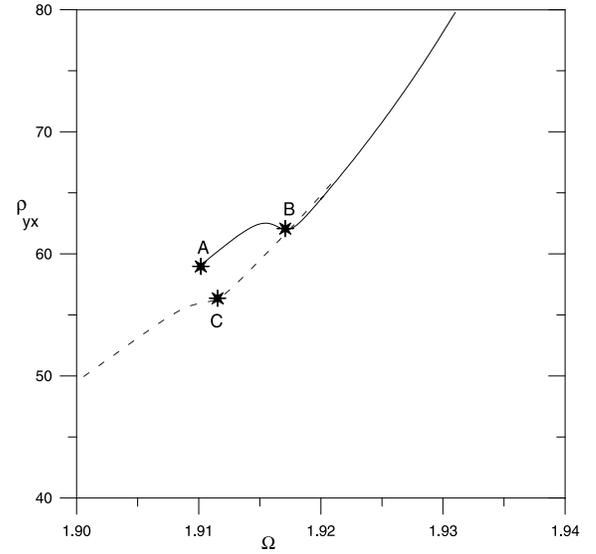}
\caption{ Hall resistivity as a function of $\Omega$. The short full line contains only $2$-body interaction ($g_2=1$) , while the long dotted one contains also $3$-body interaction ($\,g_3=2$). We consider $N=5$ and $(\bs a=(0.8,0)$. The marked points are: $A$: $(1.91,58.94)$ and $B$: $(1.9172,62.61)$ on the full line and $C$: $(1.911,56.21)$ on the long dotted line.  } 
\end{figure}

\begin{figure}[tbp]
	\centering
	 \includegraphics*[width=5.5cm]{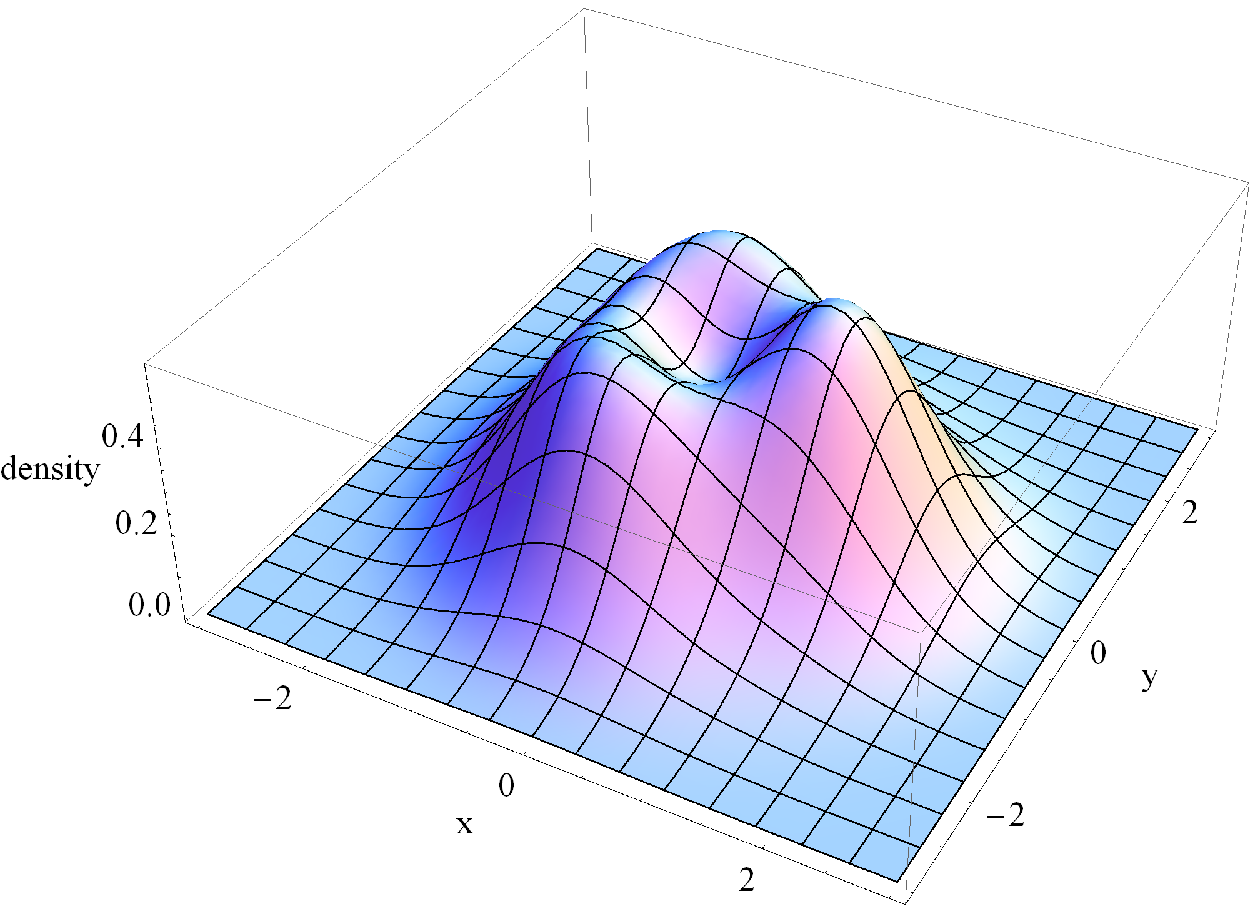}

	\caption{ 3D image of the GS density for $N=5$ at $\Omega=1.911$ (in units of $\omega_{\perp}/2$) with an impurity at $\bs a=(0.8,0)$ (in units of $\lambda_{\perp}$) and $\gamma=0.1$. We consider $g_2=1$ and $g_3=2$. } 
\end{figure}

\begin{figure}[tbp]
	\centering
	\includegraphics*[width=5.5cm]{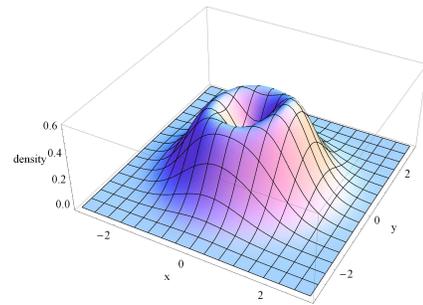}
\caption{ 3D image of the GS density for $N=5$ at $\Omega=1.911$ (in units of $\omega_{\perp}/2$) with an impurity at $\bs a=(0.8,0)$ (in units of $\lambda_{\perp}$) and $\gamma=0.1$. We consider $g_2=1$ and $g_3=0$. } 
\end{figure}

\medskip

In order to easily compare results from different $N's$, in the numerical calculation, we replace $g_2$ and $g_3$ in the interaction term, with $g_{22}=g_2\,6/N$ and $g_{33}=g_3 \,6/N$. Then, for example the same critical value $\Omega_c=\omega_{\perp}-\frac{N g}{8\pi}=1.76$ is obtained for all $N's$. $\Omega_c$ marks the place where the angular momentum jumps from $L=0$ to $L=N$,  where the first vortex is nucleated, see Figs.1 and 2.

\section{ Origin of plateaux}

In this section, we focus on the mechanism that generates plateaux of $\rho_{yx}$ within the LLL regime, namely, we refer to the fractional quantum Hall (FQH) scenario. 

The plateaux are labeled by a unique number, its filling factor, which is the manifestation of the topological nature of the associated correlated states. In the interval between subsequent plateaux, the resistivity exhibits a linear behavior on $B^*$ ($\,= \Omega\,$) given by   

\begin{equation}
\rho_{yx}\,\sim \,\frac{B^*}{n_e} 
\end{equation}
reminiscent of the classical functionality \cite{yos}.
$n_e$ is the areal density of the extended part of the system, the part that contributes to the current $\vec{j}_y$. As $B^*$ grows, the equivalent and simultaneous increase of $n_e$ is necessary to maintain the resistivity constant. Then, during certain intervals of $B^*$ transfer from localized to extended orbitals must take place, increasing $n_e$. Impurities play the role of a reservoir of particles trapping or releasing them as $B^*$ changes. 

To be more precise, we analyze the properties of the GS. As $B^*$ increases, the angular momentum of the system also increases. In the symmetric case (without impurities) where $L$ is well defined, there are abrupt changes and only certain "magic" values of $L$ are possible. For $N=4$ : $\,0-4-8\,$ and $\,12\,$, and for $N=5$: $\,0-5-8-12-15\,$ and $\,20\,$. Differently,  if some amount of asymmetry is included, the variation is soften and the expectation value of the angular momentum has all the possibilities (see Figs.1 and 2). 

\medskip

If one analyzes the occupations of the two most important natural orbitals, one can verify that there is a correspondence between the angular momentum transitions and the significant variation  of the occupations. 

\medskip

Amazingly, in the analyzed cases, there is a clear localization at the impurity of one of the natural orbitals, the first one with the largest occupation ($n_1$). Around the angular momentum transition, the decrease of $n_1$ and the increase of $n_i,\, i\geq 2$ means that transfer from localized to extended orbitals is taking place and a plateau is possible. That is to say, intervals of $B^*$ where the occupations of the natural orbitals have a significant variation, turned out to be crucial to identify the regions where transfer is possible and plateaux are expected.  

Fig.3 shows for $N=4$ the appearance of a small plateau in the region where $\,L\,$ changes from $8$ to $12$ and Fig.4 shows for $N=5$ two pseudo-plateaux around the transitions from $5$ to $8$ and from $8$ to $12$ respectively, in the symmetric case. 

It must be realized that these plateaux appear around special values of $\rho_{yx}$ that localize states of significant interest, which would be characterized by fractional filling factors in the thermodynamic limit. Without impurities these values of $\rho_{yx}$ would not be visible. This is due to the fact that the interval where $n_1$ decreases would be reduced to a point and the plateau would disappear.  Moreover, the extension of the plateau is related to the intensity of the impurity.

\section{edge excitations}

An extended way to identify the GS that results from exact diagonalization is by its overlap with a given analytical expression. However, within the FQH regime, it has been previously stressed \cite{wen1,wen2,caz2} that a more convenient and non-ambiguous way is given by its "topological order", a unique characteristic reflected in the properties of the edge excitations. This alternative becomes of special interest in our case, as in the calculation of the Hall resistivity all the spectrum is implied and not only the GS. 

\medskip

The main ingredient necessary to perform the identification is given by the number of edge excitations of the GS. References \cite{caz1,caz2} provide, for some large systems, the sequence of the number of excitations with  angular momentum $L_{GS}+m\,$ where $m \geq 0$. We resort to the results given in Ref.\cite{caz2} and identify the two states related to the two plateaux obtained for $N=5$ (see Fig.4). If the sequence of the number of the excitations from our results follows the numbers of large systems  (aside deviations due to finite size effects), for a given case considered in Ref.\cite{caz2}, then it means that with good approximation, we have found the precise vortex liquid state.

\medskip     

To count out the number of excited edge states, we proceed as follows. We analyze the interaction energies $E_{int}$ as a function of the angular momentum starting from $L_{GS}$ (see Figs.5 and 6). For each $L_{GS}+m$ we obtain a column of values. The distance between the lowest $E_{int}$ at $L_{GS}$ and the next one at the same $L_{GS}$ defines a gap. The number of $E_{int}$ for a given $m$ that lies within this gap defines the number of edge excitations for $m$. 

\medskip
 
In the case of the Laughlin state for $N=4$, this method provides the confirmation of its nature. In this case, the excited states are degenerated. We proved that this degeneracy obeys the theoretical predictions \cite{coo1}: An excited state with $L=\,N(N-1)\,+\,m\,$ is $\,p(m)\,$ times degenerated, where $p(m)$ are the partitions of $m$: the number of distinct ways $m$ can be written as a sum of smaller non-negative integers. It gives the sequence : $1-1-2-3-5-7$ for $m=0-1-2-3-4-5$, see the Table I. From our results, the plateau appears at about $\Omega\,=\,1.963$, well inside the region where the expectation value of $L_{GS}$ is close to $L=12$. Not exactly $12$ due to the presence of the small impurity that breaks the circular symmetry. 

\medskip

Table I. The Laughlin case. Number of excited states for $N=4$. The first row for an infinite system and the second row from our results. The sequence corresponds to $m=0,1,\dots,6$.

\begin{tabbing}
m=0 \= m=1 \= m=2 \= m=3 \= m=4 \= m=5 \= m=6 \kill
1 \> 1 \> 2 \> 3 \> 5 \> 7 \> 11 \\
1 \> 1 \> 2 \> 3 \> 5 \> 6 \>  8 \\
\end{tabbing}

\medskip

It must be realized that the assignation of a filling factor of $\,\nu=1/2\,$ to this state is due to the fact that the trapping potential is nearly suppressed by the strong rotation, see Eq.6. We end up with a nearly homogeneous system, for which the filling factor is well defined (it does not depend on $\bf r$). However, for other correlated states produced at lower values of the magnetic field, strong size effects prevent the association of a fractional filling factor, and the only justified assertion is that this state presents good symptoms to be identified with the correlated state with well defined filling factor in the thermodynamic limit. This is our case related to the states at the plateaux for $\,N=5\,$.

\medskip

The second plateau for $N=5$ at $\,\Omega=1.947\,$ (see Fig.4) has several properties that led us to the conclusion that it is the precursor of the correlated state with $\,\nu=2/3\,$, well approximated by the composite-fermion (CF) model. The starting requirement is the identification of $L_{GS}$. This plateau lies within the region close to the angular momentum $\,L=12\,$, one of the magic values for $\,N=5\,$, related to incompressible states. For them, the interaction energy does not change when the angular momentum is increased, the internal energy remains the same and the increase of the kinetic energy is due to the global movement of the center of mass. Once $N$ and $L$ are fixed,  then, as in the previous case, we proceed to count out the number of excited edge states (see Fig.6 and the Table II).

\medskip 
  
 Within the CF theory, the bosonic atoms are replaced by non-interacting composite particles consisting of bosonic particles with the attachment of one or more quantum of magnetic fluxes. These composite particles with fermionic statistics fill several CF LL's in a compact way. At the end, the fractional filling factor for bosons is transformed into integer filling factor for the composite particles. The relationship between the angular momenta is given by $\,L=L_{CF}+\frac{N(N-1)}{2}\,$ \cite{coo1}. In our case it gives $L_{CF}=2$ which has only one possible way to fill the CF LL's: $3$ CF on the LLL and $2$ on the first LL,  this state  is denoted as $\{3,2\}$. Similarly, for $N=6$, $\,L=20\,$ and $N=7$, $\,L=30\,$ the CF states $\{4,2\}$ and $\{5,2\}$ are obtained \cite{caz2} (in general $\{N-2,2\})$. For these three cases, the number of excitations is given in Table II. This state in the thermodynamic limit belongs to the series  $\,\nu_F\,=\,\frac{p}{2p+1}\,$ \cite{hal,jai}, the values at which fractional quantum Hall states are realized, being  $p$ the number of occupied CF LL's. For $p=2$ it gives $\,\nu_F=2/5\,$ and from the relation $\frac{1}{\nu_F}\,=\,\frac{1}{\nu_B}\,+\,1\,$, valid for homogeneous systems \cite{xie}, the filling factor for the bosonic system is $\,\nu_B=2/3\,$. 

\medskip

Table II. The state with filling factor $\nu=2/3$. The first row for an infinite system \cite{caz2}, and the next rows for $N=5$, $L_{GS}=12$; $N=6$, $L_{GS}=20$; and $N=7$, $L_{Gs}=30$, respectively. The sequence of excitations is $m=0,1,\dots,4$.
				
 \begin{tabbing}
m=0 \= m=1 \= m=2 \= m=3 \= m=4 \= m=5 \= m=6 \kill
1 \> 2 \> 5 \> 10 \> 20 \\
1 \> 2 \> 4 \> 7 \> 10 \\
1 \> 2 \> 5 \> 8 \\
1 \> 2 \> 5 \> 9  \\
\end{tabbing}

In the case of the state related to the first plateau at $\Omega=1.915$ (see Fig.4), although the overlap with the analytical expression of the Pfaffian state  \cite{coo2} is excellent (see Fig.8 commented below), the analysis of the number of edge excitations is not conclusive. 

\medskip

The two spectra obtained (not shown) for $g_2=1$, $g_3=0$ in one case and for $g_2=g_3=1$ in the other case for the first plateau, are similar to that of  Fig.6 for the second plateau. However, an important difference comes from the violation of the Kohn theorem as it is not fulfilled by some states at the upper part of the spectrum with $g_3=1$ . This is due to the fact that the center of mass and the relative variables can only be separated when the interaction depends only on differences of pairs of coordinates \cite{jac}, which is not the case for the $3$-body interaction. In spite of that, the sequence of the number of edge states is in both cases similar to the case on an infinite system, as it is shown in Table III. 
       
\medskip

Table III. The Pfaffian state. The  first row for an infinite system calculated with only $3$-body interaction and odd $N$ \cite{caz2}. For $N=5$, the second and third rows are for $g_2=g_3=1$ and $g_2=1$ $g_3=0$ respectively. The sequence corresponds to $m=0,1,\dots,4$.

 \begin{tabbing}
m=0 \= m=1 \= m=2 \= m=3 \= m=4 \= m=5 \= m=6 \kill
1 \> 2 \> 4 \> 7 \> 13 \\
1 \> 2 \> 4 \> 5 \> 8 \\
1 \> 2 \> 4 \> 6 \> 9 \\
\end{tabbing}  

Curiously, the similarity of our results with the first row is better when we do not include $3$-body interaction.
						
\medskip

We would like to make two additional comments, first, in our small system, as it will be discussed in the Section V, some amount of $2$-body interaction is necessary to have non-zero overlaps, and second, it must be realized that the differences of the number of excited states with respect to the infinite system comes from  finite size effects. This effect is clearly reduced in the Laughlin spectrum because the effective trapping potential given by $(\omega_{\perp}-\Omega)$ nearly disappears.

\medskip

\section{ Results}

\medskip

Our main result is the confirmation that we can localize the correlated vortex liquid states around the angular momentum transitions, see Figs.1 and 2. To obtain good experimental observability, it is necessary to fulfill  two conditions: a)  a small number of particles and b) that the impurity perturbs only slightly the stepwise behavior of $L_{GS}/\Omega$ in such a way that the plateaux  of the resistivity are well separated. Otherwise, for large systems, the distance between the angular momentum transitions significantly shortens, and the overlap between the wave functions of subsequent vortex liquid states would affect the observability. Namely, two ingredients are necessary: small systems and low strength of the impurities ($\gamma \ll 1$). If these conditions are met, then, within a given interval of $\Omega$ we know the number and localization of the interesting states. 

\medskip

Our important result concerns to our certainty of the presence of a precise number of correlated states below $\Omega_{max}$,  fixed by computational limitations. Moreover, we know at which values of $\Omega$ we must look for them. For $N=5$ and $\Omega_{max}\,=\,1.955$ we know that there are only two correlated states and indeed, in this case, we were able to identify them with high confidence. However, for larger values of $\Omega$ beyond $\Omega_{max}$, we can not exclude the possibility to find new plateaux not identifiable with states properly modeled using the CF approach.  

Once we have been able to generate plateaux and learned about their origin, the next step is the classification of the implied correlated states and analyze their properties. To achieve this aim, we exploit several tools. A powerful one is the overlap of the exact solution with known analytical expressions.

For the Laughlin case we have \cite{lau}
\medskip

\begin{equation}
\Psi^{Lau}(\{z_i\})\,=\, \prod_{i<j} (z_i-z_j)^2 e^{-\sum|z_i|^2/2}
\end{equation}
where $z_k=x_k+i y_k$. Or for the Pfaffian \cite{wil},

 \begin{eqnarray}
\Psi^{Pf}(\{z_i\}) & = & S\prod_{i<j \in \tau_1} (z_i-z_j)^2 e^{-\sum|z_i|^2/2} \,\,
\nonumber
\\
& \times & \prod_{k<l \in \tau_2}(z_k-z_l)^2	e^{-\sum|z_k|^2/2} 
\end{eqnarray} 
where $\tau_1$ and $\tau_2$ mean a partition of $N$. $S$ indicates that the wave function is symmetrized over all the possible partitions of $N$ particles into the subsets $\tau_1$ and $\tau_2$. 			

\medskip

Fig.7 shows for $N=4$ the modulus of the overlap, $|\langle \Psi_{analytic}|\Psi_{exact}\rangle |$ \cite{jul} between the Laughlin  state and the exact solution as a function of $\Omega$. Although the exact solution contains a small amount of anisotropy, the overlap is extremely  good, close to one along all the interval with angular momentum $L=12$. This result confirms that the implied correlated state is the Laughlin, but also indicates that the overlap is not enough to localize the plateau of the resistivity. The main reason is that in the overlap only the GS is implied,  whereas in the resistivity all the eigenstates play a role. The excited states suffer, along the interval of $L=12$, a redistribution as $\Omega$ changes producing changes in the resistivity.  One must look for the interval of $\Omega$ where $n_1$ decreases as it happens approximately between $\Omega=1.963$ and $1.967$, see Fig.3.   
     
\medskip

Fig.8 shows similar results for $N=5$. In this case, in the overlap, the analytic Pfaffian expression is used. The interval with significant values corresponds to $\langle L \rangle\,=\,8$ (see Fig.2) which is the angular momentum of the Pfaffian state for $N=5$. In general, $L=\frac{N(N-2)}{2}$ for even $N$ and $L=\frac{(N-1)^2}{2}$ for odd $N$ \cite{caz1}. For the full line  only $2$-body interaction was considered ($g_2=1$ and $g_3=0$) whereas for the plus and cross symbols, we used $g_3=1$ and $5$, respectively. Although the overlap clearly improves when $3$-body interaction is included, some amount of $2$-body component must also be considered. If only $3$-body interaction is used, although the interaction energy $E_{int}$ of the GS vanishes, meaning that it is the solution of the  $3$-body Hamiltonian, we find zero overlap with the analytic expression (Eq.12). Roncaglia {\it et $a\l$.} in Ref. \cite{ron} propose an attractive and original protocol to generate and stabilize the Pfaffian state in a  bosonic system submitted to a rotating trap. They conclude that the best way to follow their scheme is by the suppression of the $2$-body interaction. We were not able to reproduce this result and therefor we conclude that for our small number or particles, some amount of $2$-body interaction is necessary to have a significant number of particles within each subset of the partition of $N$. 

 \medskip

Next, we perform a complete analysis of the density and pair correlation function at the three  points marked in Fig.9. From this analysis, only some remarkable results are commented below. We concentrate on the first plateau where the overlap with the Pfaffian state is the best. The selected points intend to give the  following information: the point $A$ $(1.91,58.94)$ not related to a plateau, is taken as a reference; the point $B$ $(1.9172,62.61)$ on the plateau with only $2$-body interaction, and the point $C$ ($1.911,56.21)$ on the plateau with $2$ and $3$-body interactions.
\medskip

In each case, we analyze the symmetric density ($\gamma=0$), the density with the impurity, and the two-body pair correlation function $\rho^{(2)}(\vec{r}_0,\vec{r})$. In all cases, the reference position $\vec{r}_0$ is the point where the symmetric density reaches its maximum. 
\medskip

At $A$, we obtain the expected results for a non-correlated state. The symmetric density has a soft dimple at the center and exhibits a slight blown up at $\vec{a}$ when the impurity is introduced, with otherwise no any sign of spatial correlation. The pair correlation function does not uncover spatial order and at $\vec{r}_0$ the density is close to zero. 
\medskip

At $B$, the symmetric density is nearly flat with a slight maximum at the center. With the impurity, the density is blown up at $\vec{a}$ and some subtle correlated positions appear around $\vec{a}$. The function $\rho^{(2)}$  exhibits an anomalous distribution, not related to any hidden spacial correlation.  At $\vec{r}_0$ the density is significant. This point $B$ is a precursor of point $C$ as it has a subtle correlation in its density.
\medskip

Finally, point $C$ shows an interesting result which drives us to speculate about the meaning of the 3-body interaction. The symmetric density is quite flat with a slight minimum at the center, at odds with the result of point $B$. However, when the impurity is introduced, we obtain the density distribution shown in Fig.10. This figure provides some insight into the meaning of the $3$-body interaction. 

\medskip

Even though our system has five particles, the density shows only three peaks. A possible explanation is as follows: The Pfaffian wave function is a symmetric combination of all possible partitions of $N$. For a given partition of $N$ containing subsets $\tau_1$ and $\tau_2$ , in an effective way, the particles are classified into those in subset $\tau_1$ and those in subset $\tau_2$. Two particles pertaining to different subsets  do not interact with each other (see Eq.12). Assume that for this partition, particle 
${j}$ at $\tau_1$ is the one localized at the impurity and therefor, with a fixed  position. If $\tau_1$ has two particles, the second one would be localized around $j$ symmetrically. However, if $\tau_1$ has tree particles, the other two particles will accommodate far from ${j}$. This mutual repulsion results in a triangular pattern shown in Fig.10. The disappearance of the triangular pattern  when only $2$-body interaction is included provides a good support of our explanation.      

\medskip

\section{ Conclusions}

We have developed an effective way to find and, eventually classify, the correlated states of a cloud of interacting cold bosonic atoms in the fractional quantum Hall regime. Two conditions must be fulfilled to localize these sates with good observability: First, it must be a small system and second, it must be only slightly perturbed by some impurities. 
\medskip

The properties of the edge excitations were used as the most efficient tool to classify the states. Following the CF theory, we arrived to the conclusion that the second plateau is the precursor of a state of filling factor $\nu=2/3$ in the thermodynamic limit.  

\medskip
We proved that, although the overlap of the exact solution with some analytical known expressions of the GS wave functions is a powerful way to identify the state, it does not provide any insight on the localization or the extension of the plateaux of $\rho_{yx}/\Omega$. The crucial ingredient that localizes the plateaux is given by the variation of the occupation of the natural orbital localized at the impurity.
\medskip

The analysis of the GS density of the system that contains an impurity, provides new insight about the meaning of the three body interaction, for a state close to the Pfaffian.
\medskip

\section{Acknowledgments}

We are indebted to Maciej Lewenstein for his important contribution and to M.A. Garc\'{i}a-March for his comments and help. N.B. acknowledge partial financial support from the DGI (Spain) Grant No. FIS2013-41757-P. J.T. is supported by grants FIS2013-46570 and 2014-SGR-104.

\end{document}